\documentstyle[aps,multicol,epsf]{revtex}

\begin{document}

\title{Transition to Self-Organized Criticality\\ in a Rice-Pile Model}

\author{Alexei V\'{a}zquez and Oscar Sotolongo-Costa}

\address{Department of Theoretical Physics. Havana University. Havana 10400, Cuba}

\date{\today}

\maketitle

\begin{abstract}
The self-organized critical state is characterized by a power law
distribution of cluster sizes and other properties. However experiments with
sand and rice piles reveal distributions of avalanche sizes which are not
power law distributed. In the present letter a one-dimensional rice-pile
model which demonstrate that this is only an apparent contradiction is
introduced. Moreover, a robustness of the self-organized critical state is
obtained with increasing system size.
\end{abstract}


\begin{multicols}{2}

In their pioneer work \cite{bak} Bak, Tang and Wiesenfeld introduced the
notion of self-organized criticality to describe the behavior of extended
dissipative dynamical systems. They showed that such dynamical systems
naturally evolve to a critical state through a self-organization process
consisting of avalanches propagating through the system. The self-organized
critical state was then characterized by a power law distribution of cluster
sizes, a cluster being the spatial region over which a small local
perturbation propagates.

The paradigm of self-organized criticality is an idealized sandpile where
grains added to the pile dissipate their potential energy through avalanches
with no characteristic scale, where an avalanche is the energy leaving the pile. According to these definitions of cluster an avalanche, the size of a cluster is not the same as that of the avalanche, and their distribution functions will, in general, be different. Earlier experiments by Jaeger, Liu and Nagel \cite{jaeger} with the rotating drum showed a sharply peaked avalanche size distribution instead of a power law behavior. Then, they argued that these experiments indicate that avalanches in a sandpile do not behave in a self-organized critical manner. Latter experiments on sandpiles\cite{held,rosendahl,feder} and numerical simulations based on cellular automata \cite{kadanoff} showed wide range of avalanche sizes but no evident power law was observed.

More recently Frette {\em et al} \cite{frette} have performed experiments
with piles of rice obtaining different results according to the aspect ratio
of the rice grain. They found that rice-piles with elongated grains, larger
effective friction, give a power law distribution for large avalanche sizes
while less elongated grains give a faster decay. Thus, they also argued that
the experimental evidence on granular systems is in general inconsistent
with the hypothesis of self-organized criticality, ''the self-organized
criticality applies to some slowly driven granular systems but is not a
universal phenomenon''\cite{frette}.

In order to describe the effect of friction which may lead to different
behaviors Nu\~{n}es-Amaral and Lauritsen \cite{amaral} have introduced
sandpile models with stochastic toppling rules often known as rice-pile
models. Rice-pile models are closer to the experimental situation but these
precedent models do not give the rich behavior observed in experiments, from
sharply peaked to power law distributions of avalanche sizes.

Because of the aspect ratio of constituents of the pile, granularity plays an important role. This means that the slope can frequently overcome the critical angle and the ocurrence of a local avalanche will be given by certain probability; thus, the introduction of stochastic troppling rules is natural in this case. 

In the present letter a rice-pile model with an stochastic toppling rule
different to previous works is introduced. Our purpose is to have a
parameter which controls the local stability of the pile to mimic different
experimental situations. Changing this parameter a crossover to a
self-organized critical state with a power law distribution of cluster
sizes but with a distribution of avalanche sizes which is not power law
distributed is observed. 

This result may be an explanation of the apparent contradiction between experiments in sandpiles and sandpile celular automata models. In general experiments in sandpiles give non-power law distributions of avalanche sizes, thus claiming that self-organized criticallity is not universal. On the other hand sandpile models give in general power law distributions of cluster sizes characteristics of the self-organized critical state. Thus, is the absence of a power law distribution of avalanche sizes an evidence of non-universality of self-organized criticality?

Our model is defined as follows. Consider the one-dimensional system formed
by a set of $L$ sites, labeled by integers $i=1$ to $L$, with a wall at $i=0$ and an open boundary at $i=L+1$ . To each size a height $h_i$ is assigned and the local ''slope'' $z_i$

\begin{equation}
z_i=h_i-h_{i+1}-\delta h_0  \label{eq:1}
\end{equation}
is defined, $\delta h_0$ being a local slope below which the pile is very
stable. If at a certain site the local ''slope'' $z_i$ is grater than zero
then all the $z_{i}$ grains in site $i$ above height $h_{i+1}+\delta h_0$ will fall to the next site $i+1$ with certain probability $p(z_i)$. Since $z_i$ defined in \ref{eq:1} depends on the height at sites $i$ and $i+1$ when we change the height at certain site we must update the local slope of this site and that of its precedent neighbor. Note that this is a critical slope cellular automata model with a local stochastic toppling rule.

Thus, the stochastic rice-pile model is specified by the rules:

\begin{enumerate}
\item  Adding a particle: a site $i$ is selected at random with uniform
probability and $h_i$ is increased by $1$, i.e. $\{z_i\rightarrow z_i+1,\
z_{i-1}\rightarrow z_{i-1}-1\}$.

\item  Stochastic toppling rule: If $z_i>0$ then the system is updated
according to the set of rules $\{z_{i-1}\rightarrow z_{i-1}+z_i,\
z_i\rightarrow z_i-2z_i,\ z_{i+1}\rightarrow z_{i+1}+z_i\}$ with probability 
$p(z_i)$.
\end{enumerate}

Different dependences $p(z_i)$ may be defined. However, in order to
introduce a parameter able to deal with different experimental situations we
chose the Boltzman-like relation 
\begin{equation}
p(z_i)=1-\exp (-z_i/dz)  \label{eq:2}
\end{equation}
Other dependences $p=p(z/dz)$ which saturate for $z/dz \gg 1$ may be proposed, but the qualitative behavior will be the same.

When $dz=0$ we recover the one-dimensional trivial model \cite{bak1} with a
uniform distribution of cluster sizes. For $dz>0$ we have local stability if 
$0<z_i<dz$ and local instability otherwise. Hence, increasing $dz$ we may
obtain larger local slopes which is equivalent to increase the effective
friction or to decrease the energy added to the pile in an experimental
situation.

As in previous sandpile and ricepile models we assume separation of time scales. Hence, we first start with a configuration with $z_{i}=0;i=1,...,L$ and we add a grain, increase the value of $z$ at a site taken at random. Then we measure the cluster size (i.e. the number of sites over which this initial perturbation propagates) and the avalanche size (i.e. the number of grains leaving the pile). The next grain is added when the toppling process stops or an avalanche emerges.

\begin{figure}
\narrowtext
\centerline{\epsfxsize=3in \epsfbox{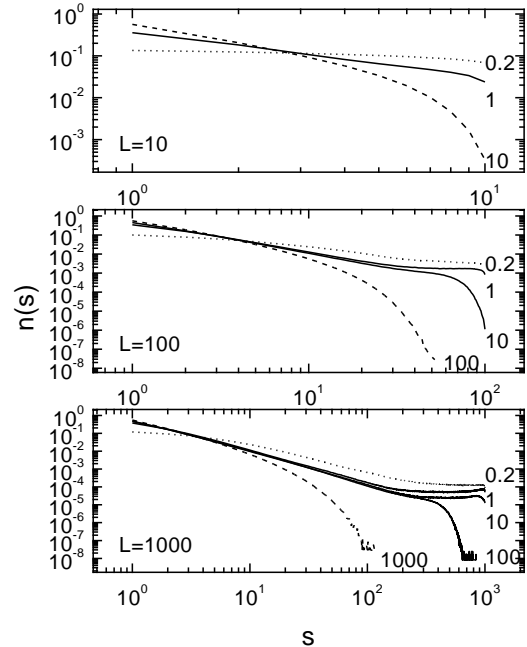}}
\caption{Distribution of cluster sizes $n(s)$ for $L=10,100,1000$ and different values of $dz$. The insets show the values of $dz$.}
\label{fig:1}
\end{figure}

The numerical experiment following the rules presented above was performed
for different values of $L$ (10, 100, 1000) and different values of $dz$
(0.2, 1, 10, 100). The results for the distribution of cluster sizes is shown in figure \ref{fig:1}. The smallest lattice $L=10$ is very sensitive to changes in $dz$. For $dz=0.2$ it exhibits an almost uniform distribution of cluster sizes corresponding to the trivial limit ($dz=0$), a power law distribution with decay exponent close to unity for $dz=1$, and a faster decay for $dz>1$. 

This picture changes with increasing lattice size. For lattice sizes $L=100$ and $L=1000$ the power law behavior is observed in a larger range of $dz$, between 1-10 and 1-100, and with exponent 1.5 and 1.6, respectively. Moreover, in this region the distribution of cluster sizes seems to be insensitive to the value of $dz$, except at the cutoff near $L$. In all cases for $dz<1$ we observe a tendency to the trivial limit which is faster the smaller the lattice size. On the other hand, for values larger than certain critical value $dz_c$ the power law behavior breaks down and a faster decay is observed.

According to these results $dz_c$ grows linearly with $L$. Therefore it is expected that for very large lattices the distribution of cluster sizes will follow a power law and will be independent of $dz$, at least for reasonable values of $dz$. On the other hand, the exponent of the power law grows with increasing lattice size. It changes in about 40 percent when $L$ goes from 10 to 100 but in less than 10 percent from 100 to 1000. Thus, it saturates to a value close to 1.6 with increasing lattice size.g

In such a case, we may say that the system is in a self-organized critical state. The system approaches, through a self-organization process, a critical point with fluctuations in all size scales characterized by a power law distribution of cluster sizes \cite{bak}.

\begin{figure}
\narrowtext
\centerline{\epsfxsize=3in \epsfbox{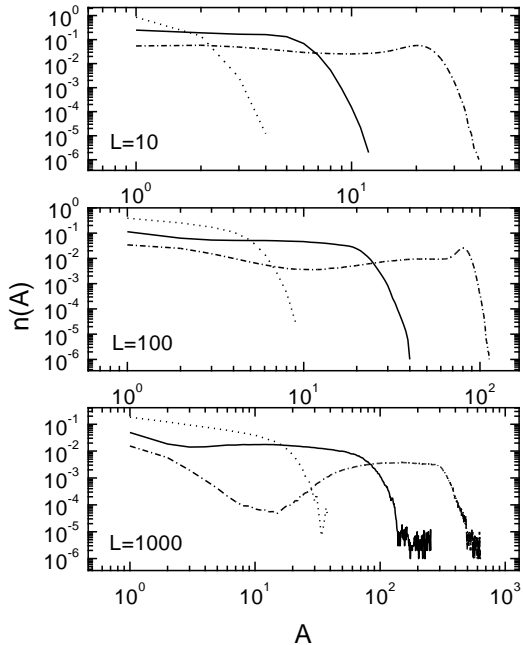}}
\caption{Distribution of avalanche sizes $n(A)$ for $L=10,100,1000$ and different values of $dz$. $dz=0.2$ dashed line, $dz=1$ continuous line and $dz=10$ dashed-dotted line.}
\label{fig:2}
\end{figure}

The distribution of avalanche sizes for $L=10$, $100$ and $1000$ is shown in
figure \ref{fig:2}. In all cases a transition from a distribution having a power law decay for small avalanche sizes, with a sharp cutoff for large sizes, to a distribution with preference for the largest avalanche sizes is observed. Besides, this transition becomes more evident the larger is the system size, while for small piles is only obtained for very large values of $dz$. This result is in agreement, at least qualitative, with the experimental evidence. The experiments by Jaeger {\em et al} \cite{jaeger}with very large piles ($L\sim 100$) showed sharply peaked distribution of avalanche sizes while the experiments by Held {\em et al} \cite{held} with smaller piles ($L\sim 10$) showed non-peaked distributions.

Hence, in general, the avalanche size is not power law distributed even if
the system is in a self-organized state with a power law distribution of
cluster sizes. The observation of distributions of avalanche sizes which are
not power law distributed does not rule out the existence of a self-organized critical state. 

From our numerical simulations we conclude that in extended dissipative
dynamical systems with relative small sizes the self-organized critical state is not universal, it only applies to some particular situations. However, increasing the system size the self-organized state becomes more and more predominant. In the thermodynamic limit the self-organized critical state is expected to be the natural state of the system.

For the first time we have introduced an stochastic toppling rule which gives different behaviors for the avalanche size distribution. In this way we have described different experimental situations, from power laws to distributions with preference for the largest avalanche sizes.

\end{multicols}

\end{document}